# Substrate induced magnetic anisotropies and magneto-optical response in YIG nanosized epitaxial films on NdGG(111)

B.B. Krichevtsov, V.E. Bursian, S.V. Gastev, A.M. Korovin, L.V. Lutsev, S.M. Suturin,

K.V. Mashkov, M.P. Volkov and N.S. Sokolov*

Ioffe Institute, 26 Politechnicheskaya str., 194021 Saint Petersburg, Russia

*Corresponding author, E-mail: nsokolov@fl.ioffe.ru

## Abstract

Nanosized $Y_3Fe_5O_{12}$ epitaxial films have been grown on $Nd_3Ga_5O_{12}$ substrates using laser molecular beam epitaxy method. Magneto-optical polar Kerr effect, ferromagnetic resonance and spin wave propagation measurements show that the stress-related anisotropy field has an opposite sign, compared to that in the YIG/GGG systems. This leads to a considerable decrease of the effective magnetization that opens a perspective to get YIG films with perpendicular magnetization for utilizing forward volume spin waves. Longitudinal magnetooptical Kerr effect magnetometry reveals a large contribution of quadratic in magnetization terms into dielectric permittivity tensor at optical frequencies. This effect strongly increases with temperature decrease and is explained by magnetization of the interface $Nd^{3+}$ ions that are exchange coupled to the $Fe^{3+}$ ions.

The nanoscale heterostructures based on yttrium iron garnet (YIG, $Y_3Fe_5O_{12}$) films attract much attention nowadays owing to the intense development of oxide spintronics and magnonics [1,2,3,4]. For the realization of effective data transmission and processing by spin wave packets, one needs high quality heterostructures with low magnetic losses at GHz frequencies. A great number of recent studies has been dedicated to investigation of epitaxial growth, static and dynamic magnetic properties of YIG layers epitaxially grown onto gadolinium gallium garnet (GGG, $Gd_3Ga_5O_{12}$) substrates. The nanosized YIG films grown by laser molecular beam epitaxy (LMBE) exhibit high crystalline quality [5,6,7,8], narrow-line ferromagnetic resonance (FMR) and low spin wave damping [9,10,11,12,13]. Assuming that the damping parameter α is linear function of the resonance frequency, in Ref. 10 it was calculated that $\alpha = 6.15\times10^{-5}$. On the other hand, the direct measurement of spin wave propagation in nanosized YIG films grown at low temperatures (700 °C) shows that the damping parameter can be even lower ($< 3.6\cdot10^{-5}$) [12].

In the absence of external magnetic field, the magnetization in YIG / GGG(111) films lies in-plane due to the demagnetizing field $H_d = 4\pi M_s \approx (1.2 – 1.5)$ kOe reinforced by the magnetic anisotropy $H_a \approx -1$ kOe ($4\pi M_{eff} = 4\pi M_s – H_a \approx 2$ kOe) induced by the magnetoelastic interactions [14,15]. In the forward volume spin wave devices the out-of-plane magnetization is necessary [16,17,18] and a strong permanent magnet is required. The way to avoid the bulky magnet, is to construct

YIG films with perpendicular magnetic anisotropy. This can be done by modification of $H_a$ by strain engineering (e.g. choosing a proper substrate [19], [20]). It was shown in Ref. 18 that in a 115 nm thick YIG film grown on neodymium gallium garnet ($Nd_3Ga_5O_{12}$, NdGG) substrate, the rhombohedral distortion is of opposite sign compared to the YIG / GGG (111) system thus reducing the magnetic field required for out-of-plane magnetization orientation.

In this paper, we investigate static and dynamic magnetic properties of considerably thinner (35 and 12nm) YIG epitaxial films grown on NdGG (111) substrates by LMBE. The magnetization reversal, ferromagnetic resonance (FMR) and spin wave propagation (SWP) in YIG / NdGG heterostructures are studied by magnetic and magneto-optical methods for differently oriented magnetic field.

The NdGG and GGG substrates were annealed before the growth (3 hours at 1000°C, ambient atmosphere) to make the surface atomically smooth. Excimer KrF Lambda Physics COMPEX 201 laser ($\lambda \approx 248$ nm) was used to ablate the YIG target at a fluence of 3.0-3.4 J/cm$^2$. The temperature of NdGG substrate was in the 500-850°C range (see Table 1, samples #1-#4). For the reference, YIG film has been also grown on GGG(111) substrate (sample #5). The growth mode is layer-by-layer as confirmed by specular spot intensity oscillations in reflection high energy electron diffraction (RHEED) pattern, Fig. 1a. RHEED pattern shown in Fig. 1b and atomic force microscopy (AFM) image, Fig. 1c, confirmed that YIG on the NdGG substrate forms epitaxial films with atomically smooth surface. High quality YIG films were recently fabricated on GGG substrates using similar growth conditions [8], [21].

| Sample | Growth temperature, °C | Thickness, nm | $4\pi M_s$ (VSM), kG | $4\pi M_{eff}$ (PMOKE), kG | $4\pi M_{eff}$ (FMR), kG | $4\pi M_{eff}$* (SWP), kG |
|---|---|---|---|---|---|---|
| #1 | 500 | 35 | 1.5 | 1.3 | [0.80 … 1.25] | 1.24 |
| #2 | 700 | 35 | 1.5 | 1.2 | [0.78 … 1.23] | 1.08 |
| #3 | 850 | 35 | 1.7 | 0.5 | [0.17 … 1.19] | 0.58 |
| #4 | 700 | 12 | 1.5 | 0.6 | [ ≤0.08 … 0.80] | 0.64 |
| #5 | 600 | 35 | - | 2.2 | 1.90 | - |

*In spin wave propagation experiments, $4\pi M_{eff}$ values were calculated using the frequency of the strongest maximum in $S_{21}$ spectra.

Table.1. Growth parameters and magnetic properties of YIG films grown on NdGGG (#1-#4) and GGG substrates (#5). Given are growth temperature, film thickness, saturation magnetization $4\pi M_s$, measured by VSM and effective magnetization $4\pi M_{eff} = 4\pi M_s - H_a$, obtained from PMOKE, FMR and SWP experiments.

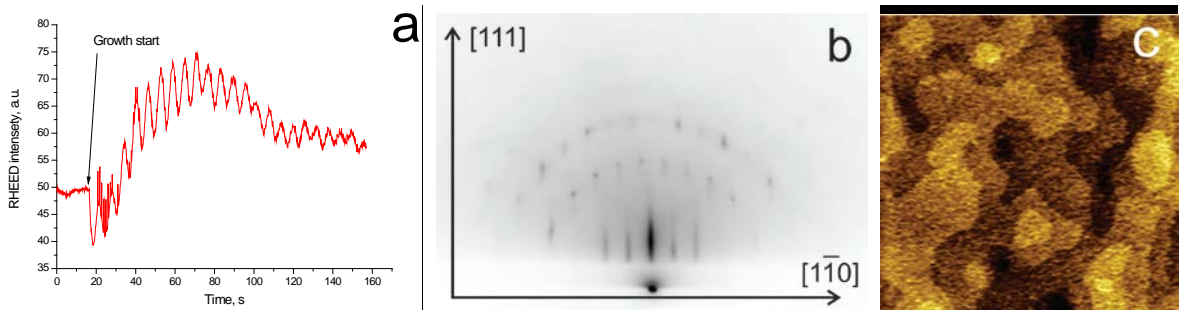

Fig.1 Characterization of growth mode and structural quality of YIG film #2. (a) RHEED specular beam intensity oscillations during the film growth, (b) RHEED pattern, (c) AFM image (3000 x 3000 x 1 nm) of the film surface.

In-plane magnetization curves were measured using Quantum design PPMS-9 vibrating sample magnetometer (VSM). The observed hysteresis loops (not shown) were narrow and had saturation magnetization of $4\pi M_s \sim 1.6$ kG (Table 1).

Fig. 2 illustrates static magnetic properties of the films. The out-of-plane magnetization curves, measured by PMOKE ($\lambda$ = 405 nm) in a series of YIG / NdGGG samples, saturate at magnetic fields $H_s$ close or less than $4\pi M_s$ value measured by VSM. This is in contrast to the earlier studied YIG / GGG system in which $H_s > 4\pi M_s$ [8,21]. From saturation field $H_s = 4\pi M_{eff} = 4\pi M_s - H_a$ we can estimate the effective magnetization $4\pi M_{eff}$ and anisotropy field $H_a$. In contrast to YIG/GGG system the sign of anisotropy field $H_a$ is positive, i.e. the induced magnetic anisotropy favors the out-of-plane magnetization. The $H_a$ may reach a large value: e.g. $H_a \approx +1$ kOe in the sample #3.

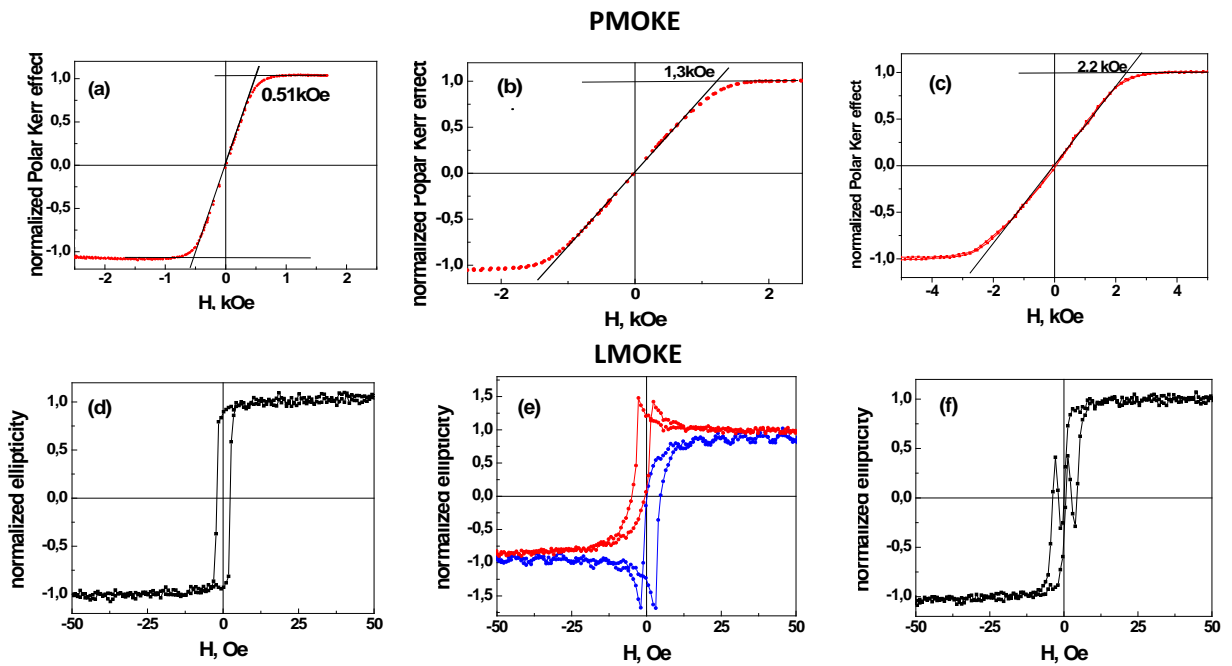

Fig.2 Static magnetic properties of YIG/NdGG heterostructures. (a,b,c) - magnetization curves measured by PMOKE (polarization plane rotation) in samples #3, #1 (YIG/NdGG) and #5 (YIG/GGG). Magnetization curves measured by LMOKE (ellipticity) in sample #2 (YIG/NdGG) for in-plane magnetic field oriented along the in-plane easy axis (d), at angle +45° (red) and -45° (blue) to the easy axis (e), and along the in-plane hard axis (f).

The LMOKE in-plane magnetization curves in YIG / NdGG films were obtained by measuring either ellipticity or polarization plane rotation at $\lambda$ = 405 nm. The azimuthal dependence of the hysteresis loop shape exhibits a 180° periodicity characteristic of the uniaxial in-plane magnetic anisotropy. Fig. 2d-f shows ellipticity measured in sample #2 with in-plane magnetic field applied at 0°, ±45°, and 90° to the easy axis (EA). With magnetic field along the EA, the hysteresis loops are narrow and rectangular similar to those observed in YIG /GGG(111) [8],[21]. When the field direction is at an angle to the EA, the loops are strongly asymmetric, Fig. 2e. With magnetic field along the hard axis (HA), peculiar jumps probably related to some modification of domain structure appear in the magnetization loops, see Fig. 2f.

The asymmetry of the hysteresis loop shape depending on the in-plane magnetic field orientation was earlier observed in Fe and Co (110) epitaxial films [22], as well as in the Fe/GaAs(001) [23] and CoFeB/MgO(001) [24] nanostructures. The loop asymmetry is caused by the $\beta_{ijkl}(\omega)M_kM_l$ terms in the dielectric permittivity tensor $\varepsilon_{ij}(\omega,\mathbf{M})$ that are quadratic in magnetization at optical frequencies. In magneto-optical experiments performed in transmission Voigt geometry, the real part of $\beta_{ijkl}$ tensor is responsible for the Cotton-Mouton effect and the imaginary part – for the magnetic linear dichroism [25]. In LMOKE measurements, with magnetization lying in-plane, the real and imaginary parts of $\beta_{ijkl}$ tensor can contribute into ellipticity and polarization rotation, correspondingly. The polarization rotation measurements carried out in sample #2 during magnetization reversal did not show any asymmetry of hysteresis loops. This indicates that the observed asymmetry of the loops measured by ellipticity originates from the real part of $\beta_{ijkl}$ tensor. The linear and quadratic terms can be separated by measuring two magnetization curves with magnetic field rotated by a positive +θ and negative -θ angle with respect to the EA. The half-sum and half-difference of these curves will give the linear and quadratic term correspondingly, Fig. 3 a,b.

We have found that the quadratic contribution considerably increases at low temperatures and becomes higher than the linear one, Fig. 3b,c. The temperature dependence of the linear terms is considerably weaker, being approximately proportional to M(T) in YIG. Temperature decrease from 300K to 150K is followed by increase of quadratic contribution by a factor of ~6. This can be hardly associated with magnetization of iron sublattices within the YIG layer because in this temperature range the square of iron total magnetization $M^2_{Fe}$ is increased only by a factor of ~1.6 [26]. Strong increase of quadratic contribution may be associated with manifestation of interface magnetization. $Nd^{3+}$ ions in NdGG substrate are in paramagnetic state and at the interface they can be magnetized by superexchange with $Fe^{3+}$ ions from tetra- or octahedral magnetic sublattices of YIG. During the magnetization reversal, the $Nd^{3+}$ ion magnetization is coupled to that of the YIG layer. Magnetization of interface $Nd^{3+}$ ions should be proportional to product of exchange field $H_e = J\mathbf{M}_{Fe}$ and paramagnetic susceptibility $\chi \sim 1/T$. Quadratic effects should follow the $T^{-2}M_{Fe}^2(T)$ dependence, which increases by a factor of ~ 6.4 with temperature decrease from 300K to 150K [26] that is close to that observed in our experiment. Note that similar strong increase of rare-earth magnetization has been found in many rare-earth iron garnets [27]. It is also worth mentioning that in the YIG / GGG system, the induced magnetization of interface $Gd^{3+}$ ions was observed at low temperature by spin polarized neutron reflectometry [28]. Distinctive features of a nanometer thick interface layer between YIG and GGG were observed by magneto-optical spectroscopy [29].

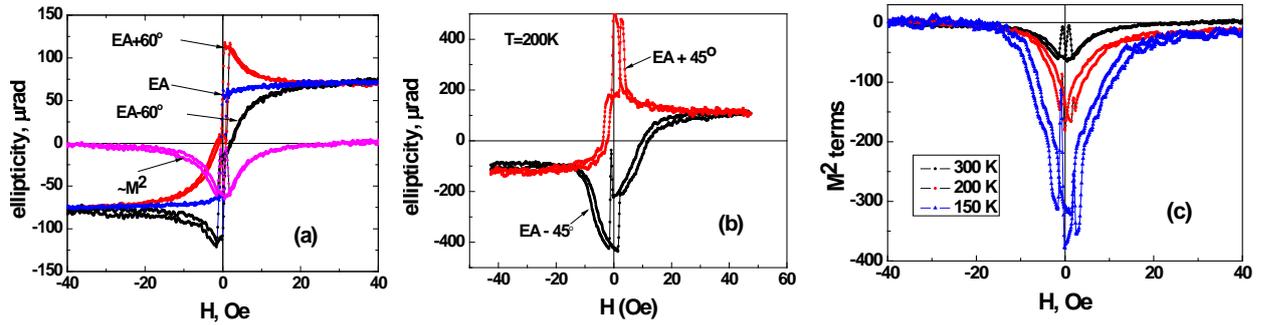

Fig. 3. LMOKE magnetization curves measured in sample #2 with magnetic field applied at different angles to EA: a) ± 60°, T = 300 K; b) ± 45°, T = 200 K. The linear and quadratic terms are shown in a) by blue and magenta curves correspondingly. Quadratic contribution into ellipticity for three temperatures is shown in (c).

The pronounced manifestation of the neodymium magnetization in the reflected light ellipticity may be related to high magnetooptical susceptibility at $\lambda = 405$ nm because this wavelength is very close to absorption lines in $Nd^{3+}$ ions in NdGG [30]. On the contrary, much smaller magnetooptic susceptibility is expected in the YIG / GGG heterostructures, because $Gd^{3+}$ is an S-ion and its absorption lines are much weaker than those of $Nd^{3+}$.

It should be noted that $Nd^{3+}$ interface ions may be magnetized by $Fe^{3+}$ ions both from tetra- or octahedral positions, which have opposite orientation of spins. Nevertheless the contribution of the quadratic magnetooptical terms $\sim \beta_{ijkl}(\omega)M_kM_l$ is the same for the opposite directions of $Nd^{3+}$ magnetic moment. For this reason, the quadratic magneto-optical effects may be pronounced even when the numbers of $Nd^{3+}$ ions with opposite magnetization orientation are comparable. In this case, the methods sensitive to the net magnetization (PNR or XMCD) will give zero output. However the quadratic magnetooptical phenomena will still be observable.

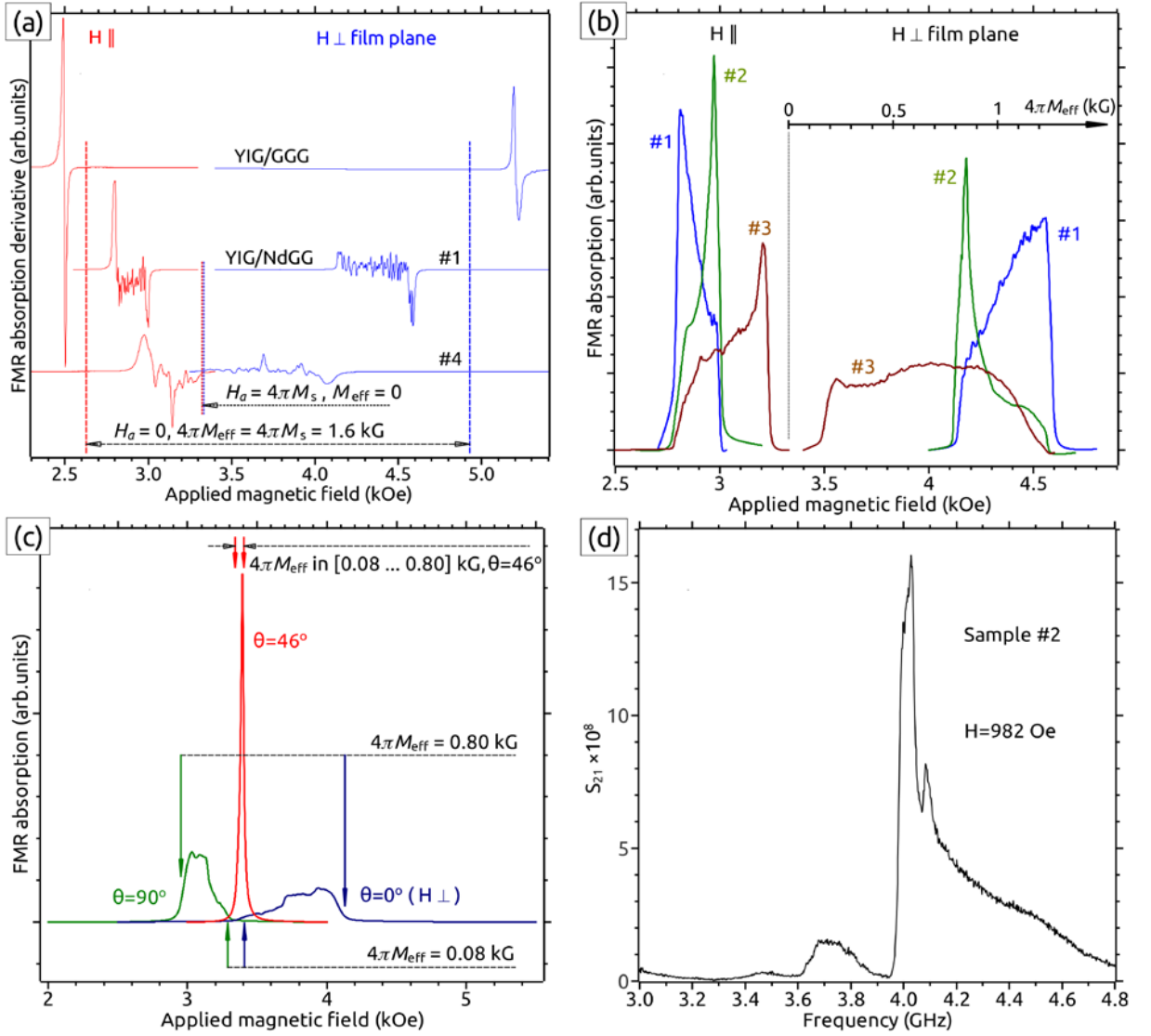

Fig.4. Dynamic magnetic properties of YIG/NdGG heterostructures. (a) Experimentally measured FMR absorption derivative spectra in samples #1 and #4, and for comparison in YIG/GGG(111) sample #5. Vertical lines show the calculated resonance fields for $4\pi M_{eff} = 0$ and $4\pi M_{eff} = 4\pi M_s$. (b) FMR absorption spectra obtained by numerical integration of experimental data. (c) FMR spectra of sample #4 for magnetic field oriented in-plane, out-of-plane and at 46° to the surface. The arrows show the calculated resonance fields for $4\pi M_{eff} = 0.8$ kG and $4\pi M_{eff} = 0.08$ kG. (d) Amplitude-frequency characteristic (scalar gain) $S_{21}$ of the spin waves propagated in the sample #2 in the in-plane magnetic field.

Dynamic magnetic properties of YIG/NdGG heterostructures are presented in Fig. 4. FMR spectra were obtained at room temperature and $F = 9.4$ GHz frequency, utilizing a conventional

ESR spectrometer with a small-amplitude modulation of the slowly scanning magnetic field, which makes spectra to appear in the form of absorption derivative (Fig. 4a). The original spectra were then numerically integrated (Fig. 4b, c) in order to make more illustrative the nature of the complicated, multi-component line structure, which is discussed below. In the figures 4b and 4c also shown are the theoretical resonance positions, calculated with Kittel formulae for the uniaxially stressed film (cubic anisotropy terms or any other in-plane anisotropy are negligible on the actual scale) [31]:

$$\left(\frac{2\pi F}{\gamma}\right)^2 = H_{res}^{in}(4\pi M_{eff} + H_{res}^{in})$$

$$\frac{2\pi F}{\gamma} = H_{res}^{out} - 4\pi M_{eff}$$

where $H_{res}^{in}$ and $H_{res}^{out}$ are resonance fields for in-plane and out-of-plane magnetic field correspondingly, $4\pi M_{eff} = 4\pi M_s - H_a$ is the effective magnetization. The $4\pi M_s$ value is taken from VSM measurements, and $\gamma/(2\pi) = 2.83$ MHz/Oe is consistent with the commonly known data [31], as well as with all spectra, presented here.

It should be first of all noted that the FMR lines, obtained for the two regular directions of magnetic field (Fig. 4a), lie closer to each other for all YIG/NdGG samples, compared with that of the YIG/GGG heterostructures. The values of $4\pi M_{eff}$, evaluated from the $H_{res}^{in}$ and $H_{res}^{out}$ resonance fields, will be evidently smaller than the average value of $4\pi M_s = (1.6\pm0.1)$ kG, as obtained from the VSM technique. This means, that $H_a$ is positive, in accordance with the PMOKE results (Table 1).

On the other hand, the FMR lines exhibit a complicated, multi-component structure within a very large total linewidth (Fig. 4 a-c). We believe, this is due to a lateral or in-depth inhomogeneity of the film. We also assume that this inhomogeneity can be characterized with a distribution of the $4\pi M_{eff}$ value. This assumption is supported experimentally by the fact that the multi-component structure shrinks to a single narrow line (Fig. 4c, red line) at some intermediate direction of magnetic field, which is consistent with the calculated angular dependences.

In this case, due to the linear relation $4\pi M_{eff} = H_{res}^{out} - 2\pi F/\gamma = H_{res}^{out} - 3.33$ kG, the integrated FMR spectra for out-of-plane magnetic field, provide the distribution density of the $4\pi M_{eff}$ parameter within the sample (Fig. 4b). The ranges of these distributions for different samples are seen from the spectra and summarized in the Table 1. Evidently, the samples #3 and #4 have regions with a very small effective magnetization.

The propagation of spin waves was measured using a couple of 2 mm x 30 μm antennas separated by 1.2 mm. A magnetic field of 982 Oe was applied in the film plane perpendicular to the spin wave propagation direction. Figure 4d presents amplitude-frequency characteristics (scalar gains) $S_{21}$ of the spin waves propagated in sample #2. The values of $4\pi M_{eff}$ in different structures calculated from SWP spectra are presented in Table 1. It is noteworthy that complicated shape of FMR absorption for sample #2 in Fig. 4b correlates with that of $S_{21}$ spectral dependence in Fig. 4d.

Summarizing the above, one can conclude, that the considerable decrease of the effective magnetization $4\pi M_{eff}$ observed in the YIG / NdGG heterostructures grown by LMBE allows us to expect that by further optimization of the growth conditions and architecture of the heterostructure, one can obtain YIG films with an out-of-plane easy magnetization axis. The manifestation of the quadratic in magnetization contribution to the reflected light ellipticity during in-plane magnetization reversal can be explained by the induced magnetization of $Nd^{3+}$ interface ions caused by superexchange interaction with $Fe^{3+}$ ions of the YIG layer. The neodymium magnetization is coupled to that of the iron in YIG, as evidenced by the anisotropy of the quadratic magnetooptical phenomena observed upon magnetization reversal and drastic increase of quadratic contribution with temperature decrease.

The LMBE growth of YIG films and experiment on spin wave propagation were supported by Russian Science Foundation (project No 17-12-01508). Magnetooptical, vibrating sample magnetometry and ferromagnetic resonance measurements were supported by Russian Foundation for Basic Research (project No 16-02-00410). The authors thank V.N. Smelov for the assistance in LMOKE measurements.